\documentclass[%
 reprint,
nofootinbib,
 amsmath,amssymb,
 aps,
]{revtex4-1}

\usepackage{graphicx}
\usepackage{dcolumn}
\usepackage{bm}

\begin{document}

\preprint{APS/123-QED}

\title{An equation of state in the limit of high densities}

\author{Ali Masoumi}
 \email{ali@cosmos.phy.tufts.edu}
\affiliation{Tufts Institute of Cosmology,
255 Robinson Hall, 212 College Ave,
Medford, MA, 02155
}%

\author{Samir D. Mathur}
 \email{mathur.16@osu.edu}
\affiliation{Department of Physics, The Ohio State University, Columbus,
OH 43210, USA
}%


\def\nn{\nonumber \\}
\def\p{\partial}
\def\t{\tilde}
\def\h{{1\over 2}}
\def\be{\begin{equation}}
\def\bea{\begin{eqnarray}}
\def\ee{\end{equation}}
\def\eea{\end{eqnarray}}
\def\b{\bigskip}


\begin{abstract}

We take string theory in a box of volume $V$, and ask for the entropy $S(E,V)$. We let $E$ exceed the value $E_{bh}$  corresponding to the largest black hole that can fit in the box. Several approaches in the past have suggested the expression $S\sim \sqrt{EV/G}$. We recall these arguments, and in particular expand on an argument that uses dualities of string theory. We require that expression for $S(E,V)$ be invariant under the T and S dualities, and that it agree with the black hole entropy when $E\sim E_{bh}$. These criteria lead to the above expression for $S$.  We note that this expression  had been obtained also  by a imposing a quite  different requirement -- that the entropy within a cosmological horizon be of order the Bekenstein entropy for a black hole of size the cosmological  horizon. We recall the earlier proposed model of a `dense gas of black holes' to model this entropy, and discuss its realization as a set of intersecting brane states. Finally we speculate that the cosmological evolution of such a phase may depart from the evolution expected from the classical Einstein equations, since the very large value of the entropy can lead to novel effects similar to the fuzzball dynamics found in black holes.

\end{abstract}

\maketitle


\section{\label{secone}Introduction}

Consider a box of volume $V$. In this box we put an energy $E$. What is the entropy 
\be
S=S(E,V)
\ee
in the limit when the energy density $\rho=E/V$ becomes large?

\begin{figure}[h]
\includegraphics[scale=.42]{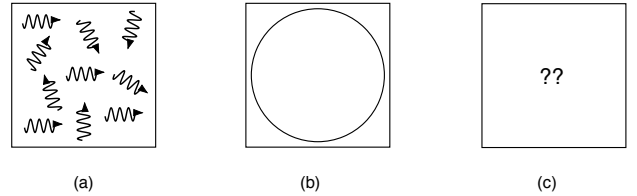}
\caption{\label{ftwo} (a) At a small value of the  energy $E$, the phase with maximal entropy is radiation. (b) At larger $E$, a black hole has more entropy; this phase continues till the size of the hole becomes of order  the size of the box. (c) We are interested in the phase where $E$ is taken to yet higher values.}
\end{figure}

  For low values of the  $E$,   we  expect the phase of matter to be radiation. This phase has  entropy $S\sim V\rho^{d/ (d+1)}$, where $d$ is the number of space dimensions (fig.\ref{ftwo}(a)). At larger $E$, we can get more entropy by forming a black hole, whose entropy is given by the expression  $S=A/(4G)$. As we increase $E$, we reach a critical value 
$E\sim E_{bh}$, where the radius of the hole $R_s$ becomes order the size $L$ of the box (fig.\ref{ftwo}(b)). 

\begin{figure}[h]
\includegraphics[scale=.32]{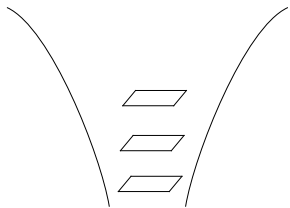}
\caption{\label{fone} A box of the same physical size at different times in an expanding cosmology. At an early enough time, the box  will contain more mass than required to make a black hole with size equal to the size of the box. }
\end{figure}

We are interested in $S(E,V)$ in the domain $E>E_{bh}$ (fig.\ref{ftwo}(c)). To see how this question makes sense, consider a flat cosmology as we follow it backwards towards the initial singularity.  In fig.\ref{fone} we depict a box-shaped region of physical volume $V\gg l_p^d$,  at different times during the evolution ($l_p$ is the planck length). The energy $E$ in the box will reach $E\sim E_{bh}$ when the density $\rho_{bh}$ in the box is still much below planck density $\rho_p$
\be
\rho_{bh}\equiv {E_{bh}\over V} \ll \rho_p
\ee
 If we push back further in time, we find $E>E_{bh}$ in our box. Einstein's equations do not constrain the value of $E$; they simply tell us that the box will be expanding at a rate given by the Friedmann equation
\be
\left ({\dot a \over a}\right )^2={16\pi G\rho \over d(d-1)}
\label{oneq}
\ee
We will let our box be in the shape of a torus $T^d$; for string theory, we have $d=9$. 
We will assume that the box size is evolving  in  the fashion (\ref{oneq}). We further assume that in spite of this expansion, it makes sense to  define an entropy $S(E,V)$. (A similar assumption is made in the standard treatment of the big bang in the radiation phase; one assumes thermal equilibrium for most computations even though the system is not, strictly   speaking, in equilibrium.)

Several approaches have suggested the answer
\be
S=C\sqrt{EV\over G}
\label{one}
\ee
Here $C$ is a constant of order unity. 
We will summarize these approaches below, but in the present paper our focus will be on using the duality properties of string theory. In \cite{sas} it was noted that the expression (\ref{one}) was invariant under the T and S dualities. We will investigate the allowed expression for $S$ by requiring such duality invariance, and asking what possible expressions for $S$ can have these invariances. More explicitly, 
we will require that $S(E,V)$ satisfy the following requirements:

\b

 (i) $S$ should be invariant under T-duality in any cycle of the torus.
 
  (ii) $S$ should be invariant under S-duality.
  
   (iii) We should get $S\sim S_{bh}$ when the box size and shape is such that $E\sim E_{bh}$ for that box. 
   
\b

With these requirements, we  argue  that we are led to the expression (\ref{one}),
in the domain
\be
\rho_{bh}\lesssim \rho\lesssim \rho_p
\label{two}
\ee
At the lower end of this range ($\rho\sim \rho_{bh}$) we will find that the expression (\ref{one}) matches onto the area entropy of the black hole $S_{bh}=A/(4G)$. At the upper end $\rho=\rho_p$ we will find that (\ref{one}) gives an entropy of one bit per unit planck volume. Thus (\ref{one}) extrapolates the Bekenstein `area entropy'  \cite{bek} to the domain (\ref{two}). Since $\rho \gtrsim \rho_{bh}$, we will say that matter is `hyper-compressed'; i.e., compressed beyond the density of the largest black hole that can fit in the box.

At this point we recall that the expression (\ref{one}) has been obtained earlier, by using quite a different line of argument. Since our box is expanding in the fashion (\ref{oneq}), the spacetime has a {\it cosmological} horizon radius $H^{-1}=\left ({\dot a \over a}\right ) ^{-1}$.  It has been suggested several times in the past that that in the very early Universe the entropy in a region of radius  $H^{-1}$ should be given by the entropy of a black hole with radius $\sim H^{-1}$; see for example \cite{fs,venezianopre}.\footnote{A more precise version of cosmological entropy bounds has been developed in terms of the entropy that can pass through light sheets \cite{fs,bousso}.}   Interestingly, this requirement gives the {\it same} expression (\ref{one}) for the entropy of a box of volume $V$. Note that  (\ref{one}) can be written as
\be
S=C\sqrt{EV\over G}=C\sqrt{\rho\over G} ~ V
\ee
so that the entropy density is
\be
s\sim \sqrt{\rho\over G}
\label{density}
\ee
From the Friedmann equation we have the radius of the cosmological horizon
\be
 H^{-1}\sim (G\rho)^{-\h}
\ee
The entropy of a black hole of radius $\sim H^{-1}$ is $S\sim H^{-(d-1)}/G$. If this is the entropy in a cosmological horizon region, then the entropy density would be
\be
s\sim {S\over H^{-d}}\sim {H\over G }\sim \sqrt{\rho\over G}
\ee
in agreement with (\ref{density}).

The expression (\ref{one}) was obtained in \cite{sas1} by arguing for a `spacetime uncertainty relation'.  
The ideas like those of \cite{fs} were explored further in \cite{bf}, where it was noted that the entropy (\ref{one}) corresponds to an equation of state $p=\rho$. A general picture was developed where horizon sized black holes coalesce as the Universe expands, so that the entropy in a  region of size $H(t)^{-1}$ remains of order the entropy of a  black hole of radius $H(t)^{-1}$.\footnote{Another approach to the entropy of the early Universe is discussed in \cite{rama}.} In \cite{brusv} the notion of a causal connection scale was used to arrive at the same equation of state (\ref{one}). In \cite{verlinde} a similar relation was argued to correspond to the Cardy formula for the density of states.

Can we find a set of matter states which  would lead to the entropy density (\ref{density})? When $\rho$ is of order the string scale, it was argued in \cite{veneziano} that such an entropy density would be obtained for a closely packed gas of string states which are at the `Horowitz-Polchinski correspondence point' \cite{hp} (i.e., at the point where the string is large enough to be at the threshold of collapsing into a black hole). We will try to flesh out this picture somewhat, by noting that states of black holes in string theory appear to be generated by sets of intersecting branes, and thus modeling the state at general $\rho$ by  closely packed sets of intersecting branes.

We have noted that an energy density $\rho$ leads to an expansion (\ref{oneq})  if we use the classical Einstein equations. But in black holes it has been found that the semiclassical dynamics expected from Einstein's equations can be altered by an `entropy-enhanced' tunneling. One finds that    the very large value of the Bekenstein entropy implies a very large measure term in the path integral. This measure term can compete with the classical Einstein action to prevent standard gravitational collapse through the horizon \cite{tunnel}. We will ask if a similar violation of semiclassical evolution  is possible in the cosmological situation.  Note that the   entropy (\ref{one}) is  large;  as in the case of the black hole, the largeness of this entropy stems from the appearance of  $G$ which brings in  the planck scale.  

The plan of this paper is as follows. In section \ref{sectwo} we check that the expression (\ref{one}) satisfies the above requirements (i)-(iii). In section \ref{secthree} we examine these requirements in more detail. 
 In section \ref{secfour} we use the equation of state $S=S(E,V)$ to write down other thermodynamic quantities for our state.  In section \ref{secfive} we present a heuristic picture of how the expression (\ref{one}) can arise from intersecting brane states; in the limit $E\sim E_{bh}$ this picture reduces to the standard  intersecting brane picture for black holes in string theory.  In section \ref{secsix} we examine the possibility of quantum effects dominating the expansion rate of the phase (\ref{one}). Section \ref{secseven} is a discussion.

\section{Outline of the derivation}\label{sectwo}

In this section we check that the expression (\ref{one}) satisfies the requitements (i)-(iii) listed in section \ref{secone}.  The check of T and S dualities was performed earlier in \cite{sas}. 

We work with 9+1 dimensional string theory. Thus the number of space dimensions is $d=9$, and the Newton constant is $G\sim l_p^8$. We set $c=\hbar=1$ throughout this paper.

We take a toroidal box with sides $L_1\dots L_9$. Consider the expression
\be
X\equiv {EV\over G}
\ee
We can see that $X$ is dimensionless; thus any function of $X$  has the correct units to be an entropy $S$. We now examine the properties of $X$.

\b

(i) The string tension is $T_s={1\over 2\pi \alpha'}$. Let us define the string length as
\be
l_s=\sqrt{\alpha'}
\ee
Under T-duality in the direction $x_1$ we get
\bea
E l_s&\rightarrow &E l_s\nn
{L_1\over 2\pi l_s}&\rightarrow& {2\pi l_s\over L_1}\nn
{L_i\over l_s}&\rightarrow& {L_i\over l_s}, ~~~i=2, \dots 9\nn
g&\rightarrow& g{2\pi l_s\over L_1}
\label{tduality}
\eea
Here $g$ is the string coupling, and we note that  Newton's constant is given by $G=8\pi^6 g^2l_s^8$. We thus find
\be
X={EL_1 L_2\dots L_9\over 8\pi^6 g^2l_s^8}~\rightarrow~ {E ({(2\pi l_s)^2\over L_1})L_2\dots L_9\over 8\pi^6 g^2 ({(2\pi l_s)^2\over L_1^2}) l_s^8}=X
\ee
so that $X$ is invariant under T-duality on any cycle of the torus.

\b

(ii) Under S-duality, any quantity remains invariant if it is expressed in planck units. We define the planck length as $G=l_p^8$, and the planck mass as $m_p=1/l_p$. We see that we can write $X$ entirely in planck units
\be
X={EV\over G}={(El_p)(V l_p^{-9})}
\ee
so that $X\rightarrow X$ under S-duality. 

\b

(iii) Let all sides of the torus be equal: $L_i=L$. Consider the energy $E=E_{bh}$ for which the radius of the black hole would be $R_s\sim L$. In 9+1 dimensions, the metric of the Schwarzschild hole has the form
\be
ds^2=-(1-{\alpha G M\over r^7})dt^2+{dr^2\over (1-{\alpha GM\over r^7})}+r^2 d\Omega_8^2
\ee
where $\alpha$ is a constant of order unity. Thus the horizon radius is $
R_s\sim (GE_{bh})^{1\over 7} $. Setting $R_s\sim L$, we get
\be
 E_{bh}\sim {L^7\over G}
 \label{three}
\ee
At this energy
\be
X={E_{bh}V\over G}\sim {L^{16}\over G^2}
\ee
But the black hole entropy is
\be
S_{bh}\sim {A\over G}\sim {L^8\over G}
\ee
Thus at the energy (\ref{three}) we find 
\be
X^\h\sim S_{bh}
\ee

\b

To summarize, if we take
\be
S\sim X^\h= \sqrt{EV\over G}
\ee
then this entropy would be invariant under T and S dualities, and would agree with the entropy of the black hole at the lower end of the domain (\ref{two}).

\section{A more detailed analysis}\label{secthree}

In this section we examine the above derivation of (\ref{one}) in more detail. First we explain in more detail what we mean by requiring that our expression for $S$ be invariant under T and S dualities. Then we explore the constraints our requirements impose of different possible expressions for the entropy. 

\subsection{Manifest invariance under duality}\label{secmono}

 String theory is characterized by a `string length' $l_s\sim\sqrt{\alpha'}$. One of the models of the early Universe is the `string gas' \cite{bv}. The entropy of the string gas has the form
\be
S_{sg}=C_1 (E l_s) + C_2 (V l_s^{-9})
\label{ssg}
\ee
where $C_1, C_2$ are dimensionless constants.\footnote{Brane gases have a similar entropy \cite{branegases}.}
Thus the string scale $l_s$ appears explicitly in this expression. For this reason, $S_{sg}$ is not `manifestly'  invariant under S-duality. Under S-duality, the elementary string is replaced by the D-string, whose tension is $T_D=T_s/g$. We may define the `D-string length' $l_d$ analogous to how we defined the string length $l_s$
\be
T_d={1\over g}{1\over 2\pi \alpha'}\equiv {1\over  2\pi l_d^2}, ~~~\rightarrow ~~~ l_d=g^{\h} l_s
\ee
Under S-duality we get $l_s\rightarrow l_d$, and we see that the expression (\ref{ssg}) for $S_{sg}$ is not invariant.

Let us clarify here what we mean by the phrase ``$S_{sg}$ is not manifestly S-duality invariant". Suppose the string coupling is weak: $g\ll 1$. Let us  place an energy $E$ in our box, with the value of $E$ being such that the phase  we get is the string gas.  For small $g$, the states of the elementary string are lighter than the states of the D-string. Thus we expect that the excitations in our box would consist  predominantly of string states, and not of D-string states. Counting these string states, would lead to the expression $S_{sg}$, which would be a correct result in string theory (for this avlue of $E$). But S-duality is an exact symmetry of string theory. So should the result (\ref{ssg}) for the entropy not be automatically S-duality invariant?

The situation is as follows. If we perform an S-duality, the coupling $g$ gets replaced by $1/g$, and the D-string becomes lighter than the elementary string. The string length $l_s$ in (\ref{ssg}) then gets replaced by the D-string length $l_d$, and with this replacement the expression $S_{sg}$ again gives the correct entropy (in our chosen energy range). 

Thus S-duality is indeed respected by the theory, but the {\it expression} $S_{sg}$ is not left invariant under this duality. Thus we say that $S_{sg}$ is not `manifestly invariant' under S-duality. In the present paper, on the other hand, we are looking for an expression $S(E,V)$ that {\it would} be left invariant under the dualities. The motivation for this requirement comes from black holes. The entropy of a black hole $S_{bh}$ is given in terms of the planck length $l_p\sim g^{1\over 4} l_s$, which is invariant under S-duality; there is no explicit appearance of the string length $l_s$ in the expression for $S_{bh}$.  In \cite{sv} the black hole was made from sets of D-branes. The tensions of these branes involved $l_s$ and the coupling $g$, but  in the overall expression for the entropy these variables appeared in a particular combination which can be written in terms of  the planck length alone. In our present approach we are conjecturing that as we push deeper into the domain $E>E_{bh}$ the expression for $S(E,V)$ will retain the property that it be manifestly invariant under S,T dualities.

To begin our discussion, let us assume that $S$ depends on  $l_s$ and $g$ only through the combination $l_p$. Suppose we assume further that $S$ depends on the volume $V$ of our torus, and not on its shape. Then we would have $S=S(E,V,G)$. Let us also assume for the moment that  $S$ was in the form of  monomial
\be
{E^a V^b\over G^c}
\ee
(We will consider more complicated forms a little later.) Since $S$ has no units, we must have
\be
-a+9b-8c=0
\label{five}
\ee
The T-duality rules (\ref{tduality}) give
\be
b-c=0
\label{six}
\ee
Equations (\ref{five}) and (\ref{six}) give
\be
a=b=c
\ee
so we are forced to the form 
\be
S\sim \left ({EV\over G}\right )^q
\ee
for some power $q$. Matching onto $S_{bh}$ at $E=E_{bh}$ as before, we find $q=\h$. Thus under the above assumptions, we see that (\ref{one}) is the only expression that satisfies our requirements.

\subsection{Examining different shapes of the torus}\label{secshape}

 In the above discussion we have assumed that the parameters of the torus enter into the expression of $S$ only through the overall volume $V$. But it is possible in principle that the expression for $S$ depends on the ratios of the sides $L_i/L_j$. To examine this possibility, we let $d$ of the space directions have a  length $l$ while the remaining $9-d$ directions have a  length $L$. We take
 \be
 L\gg l
 \label{shape}
 \ee
 and ask if our criteria imposed on $S$ suggest an answer different from (\ref{one}). 
 
 The requirements  (i) of T-duality and (ii) of S-duality remain the same as before, and thus imply no change to (\ref{one}). The requirement  (iii) that $S\sim S_{bh}$ at $E\sim E_{bh}$ needs to be re-examined however, because the black hole at energy $E_{bh}$ will be forced to a different shape. Under the condition (\ref{shape}), we should treat the $d$ small directions as compact internal directions, so that we really get a hole in $9-d$ noncompact space dimensions. (In other words, the black hole becomes  a `black string' extending along the $d$ small directions, so that the horizon now has a topology $S^{8-d}\times T^d$.) The metric for such hole has the form
\be
ds^2=-(1-{\alpha_d G_{10-d} M\over r^{7-d}})dt^2+{dr^2\over (1-{\alpha_d G_{10-d}M\over r^{7-d}})}+r^2 d\Omega_{8-d}^2
\ee
 where $\alpha_d$ is a constant of order unity and 
 \be
 G_{10-d}={G\over l^d}
 \ee
 The Schwarzschild radius is now $R_s\sim (G_{10-d}E_{bh})^{1\over 7-d}$. Setting $R_s\sim L$ we get
\be
 E_{bh}\sim {L^{7-d}\over G_{10-d}}={L^{7-d}l^d\over G}
 \label{threeq}
\ee
 and
 \be
 S_{bh}\sim {L^{8-d}\over G_{10-d}}\sim {L^{8-d} l^d\over G}
 \ee
 We see that 
 \be
  {E_{bh}V\over G}  \sim \left ( {L^{7-d} l^d\over G}\right ) (L^{9-d} l^d) {1\over G}=\left ( { L^{16-2d} l^{2d}\over G^2}\right )
  \ee
  and thus at $E=E_{bh}$
  \be
  S\sim \sqrt{EV\over G}\sim S_{bh}
  \ee
  Thus we find that the expression (\ref{one}) does not need to be modified for very asymmetrical shapes of the torus, and we conjecture that the parameters of the torus enter only through the volume $V$ and not through the moduli $L_i/L_j$.
  
  \subsection{The expression in different dimensions}
  
  We have used 9+1 dimensional string theory for our analysis, since this choice makes it easy to see the effect of T and S dualities. It is known however that the full structure of string theory is best seen through M-theory, which lives in 10+1 dimensions. The extra space direction, usually called $x_{11}$,  has a length $L_{11}$ that depends on the coupling constant $g$ of string theory. Since $g$ does not appear explicitly in (\ref{one}), one may wonder if this expression for entropy is indeed symmetrical in all 10 space dimensions of M-theory. The 10-dimensional Newton constant $G$ is related to the 11-dimensional Newton constant $G_{11}$ by
  \be
  G={G_{11}\over L_{11}}
  \ee
  Thus
  \be
  \sqrt{EV\over G}=\sqrt{E V L_{11}\over G_{11}}=\sqrt{E V_{11}\over G_{11}}
  \label{eld}
  \ee
  where $V_{11}=VL_{11}$ is the volume of the 10-dimensional spatial box which now includes the direction $x_{11}$. Thus our expression (\ref{one}) is indeed unchanged when viewed as an expression in M-theory.
  
  Similarly, we may regard $d$ of our space dimensions as `internal' directions on which we dimensionally reduce. Let the volume of these directions be $V_c$. Then the Newton constant for the remaining directions is
  \be
  G_{10-d}={G\over V_c}
  \ee
  and we can write the expression for the entropy in terms of the non-internal quantities only 
  \be
  \sqrt{EV\over G}=\sqrt{EV/V_c\over G/V_c}=\sqrt{EV_{9-d}\over G_{10-d}}
  \ee
  We again see that the expression (\ref{one}) remains valid for the dimensionally reduced theory.
  
 Put another way, the expression (\ref{density}) for the entropy density has the property that even if we regard some of the directions as small `internal' directions, we can write
 \be
 s=C \sqrt{\rho\over G}
 \label{eqstate}
 \ee
 where $s$, $\rho$ and $G$ are quantities that are defined using the non-internal directions  alone.
  
  \subsection{Eliminating more complicated expressions}
  
  In section \ref{secmono} we have seen that if we take the ansatz for $S$ to be a single monomial in $E,V, G$, then we are led to (\ref{one}) as the only possibility. But one might imagine a more complicated expression which involves terms with different powers of these variables, such that the overall sum is invariant under T and S dualities. Based on the discussion above in section \ref{secshape}, we assume that the $L_i$ appear in our expression only through the volume $V$. Let us first write $S$ as a sum of terms with different powers of our variables
  \be
  S=\sum_{\alpha,\beta,\gamma} C_{\alpha\beta\gamma} (El_p)^\alpha ( Vl_p^{-9})^\beta g^\gamma
  \label{series}
  \ee
  where we have used the planck length to make dimensionless quantities.
 Under S-duality we have
 \be
  (El_p)^\alpha ( Vl_p^{-9})^\beta g^\gamma\rightarrow (El_p)^\alpha ( Vl_p^{-9})^\beta g^{-\gamma}
  \ee
  so that we need
  \be
  C_{\alpha\beta\gamma}=C_{\alpha\beta,-\gamma}
  \label{equality}
  \ee
  To see the effect of T-duality we write our terms using the string scale $l_s\sim g^{-{1/4}} l_p$
  \be
    (El_p)^\alpha ( Vl_p^{-9})^\beta g^\gamma \sim   (El_s g^{1/4})^\alpha ( Vl_s^{-9} g^{-9/4})^\beta g^\gamma
    \ee
    Under a T-duality in the direction $x_1$, we get
    \bea
     &&(El_s g^{1\over 4})^\alpha ( Vl_s^{-9} g^{-{9\over 4}})^\beta g^\gamma \rightarrow \nn
     && \left [  (El_s g^{1\over 4})^\alpha ({2\pi l_s\over L_1})^{\alpha\over 4}\right ] \left [  ( Vl_s^{-9} g^{-{9\over 4}})^\beta  ({2\pi l_s\over L_1})^{-{\beta\over 4}}\right ] \left [ g^\gamma ({2\pi l_s\over L_1})^\gamma\right ]\nn
     &&=(El_s g^{1\over4})^\alpha ( Vl_s^{-9} g^{-{9\over4}})^\beta g^\gamma ({l_s\over L_1})^{{\alpha-\beta\over 4}+\gamma}\nn
     \eea
   We have assumed that the lengths $L_i$ appear in our expression only through the overall volume $V=\prod L_i$, so we need
   \be
     {\alpha-\beta\over 4}+\gamma=0
     \label{abg}
     \ee
     for each term in (\ref{series}). But from (\ref{equality}) we see that this is possible only if all the terms in the series have $\gamma=0$. Then (\ref{abg}) gives $\alpha=\beta$, and we see that $S$ is a function of $EVl_p^{-8}={EV\over G}$. 
     
     Note that in our physical problem have $S\gg 1$. Matching onto $S_{bek}$ at $E=E_{bh}$ then gives (\ref{one}) at leading order. But we are still allowed  to  add lower powers of ${EV\over G}$; for example we could have
     \be
     S=C\sqrt{EV\over G}[1+\alpha_1 \log \left ( {EV\over G} \right )+\alpha_2 ({EV\over G})^{-\h}+\dots]
     \ee
     We are interested only in the leading order expression for the entropy, so we will work with (\ref{one}).

  \section{Thermodynamic properties}\label{secfour}
  
Let us compute the values of different thermodynamic quantities that follow from the equation of state
  \be
  S=C\sqrt{EV\over G}
  \label{oneqq}
  \ee
  The first law of thermodynamics gives
  \be
  TdS=dE+pdV
  \ee
  Thus
  \be
  T=\left ( {\p S\over \p E}\right ) _V^{-1}={2\over C} \sqrt{EG\over V}
  \ee
  \be
  p=T\left ( {\p S\over \p V }\right ) _E = {E\over V}=\rho
  \label{eos}
  \ee
  Writing $p=w \rho$ we see that 
  \be
  w=1
  \label{weo}
  \ee
 This fact was noted earlier in \cite{sas1,bf}, and a detailed dynamics was conjectured for such an equation of state in \cite{bfm}.  

  Note that the speed of sound is given by
  \be
  v=\left ( {\p  p\over \p \rho} \right )_s^\h=1
  \ee
  Thus compression waves in this phase travel at the speed of light, mimicking a massless scalar.
  
  \section{A pictorial model}\label{secfive}
  
  The entropy (\ref{one}) was obtained in \cite{bf} from a model where the Universe was filled with a closely spaced gas of black holes. We first reproduce this estimate. Then we conjecture that the black holes could be replaced by sets of intersecting branes, extending the model  of \cite{veneziano} where the black holes were replaced by  states of the elementary string for the case when $\rho$ was string scale.

  \subsection{The entropy of a black hole gas}\label{sechole}
  
  Consider a gas of black holes, where the holes are `closely spaced'; i.e., the separation between holes not much more than the size of the holes. We work in 9+1 dimensional string theory, and let the spacelike directions be a toroidal box $T^9$. 
   
  Let the torus $T^9$ have volume $V$. Let each black hole have radius $R_s$. The number of holes is then
\be
N_{hole}\sim \left ( {V\over R_s^9}\right )
\ee
The entropy of each hole is
\be
S_{hole}\sim {R_s^8\over G}
\ee
Thus the total entropy is
\be
S\sim N_{hole} S_{hole}\sim {V\over R_s G}
\label{entropylattice}
\ee
We see that we can make $S$ as big as we want by making $R_s$ small enough. In particular,  the entropy of such configurations can exceed the entropy given by the surface area of the box. The energy of each hole is
\be
E_{hole}\sim {R_s^7\over G}
\ee
Thus the total energy is
\be
E\sim N_{hole} E_{hole}\sim {V\over R_s^2 G}
\label{elattice}
\ee
From this expression we have
\be
R_s\sim \left ( {V\over EG}\right )^\h
\ee
Substituting this in (\ref{entropylattice}) we find
\be
S\sim {1\over R_s} {V\over G}\sim \left ( {V\over EG}\right )^{-\h}{V\over G} \sim \sqrt{EV\over G}
\ee
which agrees with (\ref{one}). 

The expression (\ref{elattice}) for the energy $E$ in our box increases as we take $R_s$ to smaller values. It appears reasonable however to require
\be 
R_s\gtrsim l_p
\ee
since we do not expect black holes with size smaller than planck scale. The highest entropy and energy are then obtained for $R_s\sim l_p$, with values
\be
S_{max}\sim {V\over l_p G}\sim {V\over l_p^9}
\ee
and
\be
E_{max}\sim  {V\over l_p^2 G}\sim {V\over l_p^9}\, m_p
\ee
We see that $E_{max}$ corresponds to planck density (order planck mass per unit planck volume), while $S_{max}$ corresponds to  an entropy of order one bit per unit planck volume. 

Recall that the lowest energy $E\sim E_{bh}$ that we have considered corresponds to having just one black hole with radius of the order of our box size. As we increase $E$ above this value, the configuration splits into many black holes, till at the value $E_{max}$ we have planck energy density and planck entropy density. Thus as we traverse the range
\be
\rho_{bh}\lesssim \rho\lesssim \rho_p
\label{twoq}
\ee
the  entropy expression $S\sim \sqrt{EV\over G}$ goes from an  entropy given by the surface area in planck units to an entropy given by the volume in planck units. 

\subsection{Replacing the black holes by string states}

   We have noted that the entropy density (\ref{density}) was obtained in \cite{bf} from a model where space was filled by a set of closely spaced black holes, with the size of each hole being of order the Hubble radius.  In \cite{veneziano} a model was proposed where string states would give the required entropy. The idea was to consider highly excited states of the elementary string, at the coupling where they are about to collapse into  a black hole under their own gravity. This coupling is called the Horowitz-Polchinski correspondence point \cite{hp}, and at this point the mass and entropy can be matched, upto factors of order unity, to the mass and entropy of small black holes. It was noted in \cite{veneziano} that If we take a closely spaced lattice of such string states, then the energy density $\rho$ is string scale, and the entropy density of this lattice agrees with (\ref{density}) for this particular value of $\rho$.
   
   Such states of the elementary string correspond to what are called `small black holes' in string theory, where the title `small' refers to the fact that the radius of the hole   is of order the string length $l_s$. To understand the states of  black holes with larger radii in string theory, one has to use  other elementary objects of the theory like branes. Black holes are somewhat esoteric objects, possessing a horizon and a singularity. We are interested in see if we can replace the black holes in the description of \cite{bf} by objects that we can understand in more traditional terms.
   
   In string theory we have learnt that there is a useful  `approximate' picture of black hole microstates that is obtained in terms of intersecting branes. More precisely, we can count the number of configurations of such intersecting branes, and thereby reproduce the black hole entropy. We will review the relevant results below. We will see in the next section that the gravitational solution corresponding to these intersecting branes is a `fuzzball', which has no horizon or singularity. But for the purposes of this  section, we can just regard the intersecting branes as a generalization  of the  string states  considered in \cite{veneziano} which allows us to obtain the entropy density (\ref{density}) at any  energy density $\rho$.

        \subsection{Black holes in string theory}

Consider radiation in $d$ space dimensions, in a fixed volume $V$, with energy $E$.  The entropy increases with $E$ as $S\sim E^\alpha$, where 
\be
\alpha={d\over d+1}<1
\ee
We have $T=({dS\over dE})^{-1}\sim E^{1-\alpha}$, and the specific heat is 
\be
c_v=({dE\over dT})\sim {1\over 1-\alpha} E^\alpha
\ee
Note that to get $c_v>0$ we need $\alpha<1$. In  string gas, we have $S\sim E$, and $c_v\rightarrow \infty$. Schwarzschild black holes on the other hand have $S\sim E^\alpha$ with $\alpha>1$; for example in 3+1 dimensions we have $S\sim E^2$. The specific heat is this situation is {\it negative}. It is a challenge for any microscopic model to reproduce this behavior of $S$, since any system described by  a canonical ensemble partition function $Z$ has a {\it positive} specific heat. To describe the black hole we need a non-equilibrium system -- one with a large number of metastable states, and the number of such states should grow rapidly with energy.

\begin{figure}[h]
\includegraphics[scale=.52]{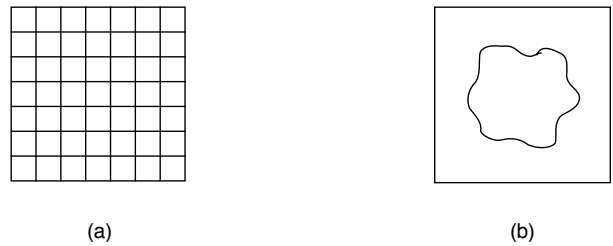}
\caption{\label{fthree} (a) Branes wrapped on different cycles have  intersection points; the number $n_{int}$ of such intersections determines the entropy as $S\sim \sqrt{n_{int}}$. (b) Branes and anti-branes can join up to make effective local objects that do not wrap all the way around the cycles of the torus; in the case depicted, the string winding-antiwinding
  and momentum-antimomentum  modes join up to make an effective string loop. }
\end{figure}

 Let us first consider extremal holes. These holes have positive specific heat, but still manifest the behavior $S\sim E^\alpha$ with $\alpha>1$. To get a hole in 3+1 noncompact directions we can compactify 6 directions $y_1\dots y_6$ on a torus $T^6$. We wrap $n_1$  D3 branes on the cycle $(y_1y_2y_3)\equiv (123)$, $n_2$ D3 branes on the cycle $(145)$, $n_3$ D3 branes on $(246)$ and $n_4$ D3 branes on $(356)$. The number of points where branes of all 4 types intersect is $n_{int}=n_1n_2n_3n_4$ (fig.\ref{fthree}(a)).   The entropy of such configurations is then given by \cite{sv,fourc}
 \be
 S\sim \sqrt{n_{int}}=\sqrt{n_1n_2n_3n_4}
 \ee
 Let each of the $n_i$ be large and of the same order $n_i\sim n$. Then for a given size of the torus $T^6$ we have
 \be
 E\sim n, ~~~S\sim n^2, ~~~S\sim E^2
 \ee
 which agrees with the behavior of the entropy of the 3+1 dimensional extremal hole. Done carefully, this computation reproduces the correct numerical factor as well, so we get $S=S_{bh}={A\over 4G}$. 
 
 A homogeneous cosmology is expected to be charge neutral, since the flux lines have no place to escape. Thus we now recall the results on nonextremal holes in string theory.  The first extremal black hole to be studied was the D1D5P hole in 4+1 noncompact dimensions. There are three charges: D1 branes, D5 branes, and momentum modes. The numbers of these charges are denoted by  $n_1, n_5, n_p$ respectively. the entropy is  \cite{sv} 
 \be
 S=2\pi \sqrt{n_1n_5n_p}
 \ee
 If we let $n_1, n_5 \gg n_p$, then the entropy of the near extremal hole is given by `momentum-antimomentum pairs' \cite{callanmalda}
 \be
 S=2\pi \sqrt{n_1n_5}(\sqrt{n_p}+\sqrt{{\bar n}_p})
 \ee
 where $\bar n_p$ gives the number of anti-momentum modes.
 If we let $n_5 \gg n_1, n_p$ the entropy is reproduced by the expression
 \be
 S=2\pi \sqrt{n_5}(\sqrt{n_1}+\sqrt{{\bar n}_1})(\sqrt{n_p}+\sqrt{{\bar n}_p}) 
 \ee
 so we have D1 branes and anti-D1 branes, as well as momentum and antimomentum modes. What is remarkable is that the entropy $S$ can also be exactly reproduced in this case by the configuration depicted in fig.\ref{fthree}(b). Here the entropy comes from the states of   an effective string of tension $T_{D}/n_5$, living inside the D5 branes, where $T_{D}$ is the tension of the D1 brane \cite{malda5}. If we think of the left side of this loop as `winding' along the cycle in the vertical direction, then we can think of the right side of the loop as `antiwinding' along this cycle. Thus branes and antibranes can join up to make localized objects. (The momentum and antimomentum modes are similarly given by excitations of the string running clockwise and anticlockwise around this string loop.) The lesson we extract from this picture is the branes and antibranes that arise in nonextremal configurations can form local compact configurations that need not extend all the way across the torus. 
 
 If all charges are comparable ($n_1\sim n_5\sim n_p$) and we have an arbitrary amount of nonextremality, then the entropy is reproduced by the expression \cite{hms}
 \be
 S=2\pi (\sqrt{n_5}+\sqrt{{\bar n}_5})(\sqrt{n_1}+\sqrt{{\bar n}_1})(\sqrt{n_p}+\sqrt{{\bar n}_p}) 
  \label{threec}
 \ee
 In particular this reproduces exactly the entropy of the Schwarzschild hole in 4+1 dimensions when all net charges are set to zero.
 
 From the above discussion, we extract the following lessons:
 
(a) String theory has highly entropic configurations given by intersecting branes.

(b) Neutral configurations can be obtained by taking both   branes and antibranes.

(c) Branes and antibranes can join up into compact localized excitations that do not have to wrap all the way around the cycles of the bounding torus.

\subsection{The entropy in a box}

In the black hole states  discussed above, some directions were compactified to a torus.  But other directions were noncompact, where the   configuration can expand to take the shape suggested  by the classical black hole horizon. In our present cosmological problem such is not the case; we have compatified all space directions, and allowed enough energy $E$ so that the Schwarzschild radius $R_s(E)$ corresponding to this energy  is bigger than the size of our box. What do we expect for the entropy $S$ in this situation?

It is sometimes said that a single black hole is the configuration with largest entropy for a given energy; breaking it up into many smaller holes will reduce the overall entropy. One may conclude from this that the maximal possible entropy in our box can be no more than the surface area of the box. 

But as we have seen in section \ref{sechole}, such a conclusion is not correct. The above statement about black hole entropy holds only when enough volume  is available  to allow a single large black hole to exist in that volume. If we limit our volume to a give value $V$, and allow sufficiently large energy $E$, then a set of small black holes can give more entropy that predicted by the surface area of the box. 

\begin{figure}[h]
\includegraphics[scale=.52]{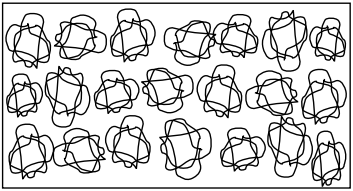}
\caption{\label{ffour} A pictorial depiction of the configuraions that reproduce the entropy (\ref{one}). Clusters of intersecting branes give the entropy of order the black hole entropy for each cluster. The overall entropy is then the sum of these entropies.}
\end{figure}

Let us now describe the heuristic brane model that will reproduce the entropy (\ref{one}). In line with the lessons (a)-(c) of the above section, we imagine the brane configuration depicted in fig.\ref{ffour}. We have intersecting branes of compact form, producing local structures that tile up to fill up torus.  The entropy of each intersecting brane set should be similar to the entropy of a black hole, so we can estimate the overall entropy by considering a lattice of black holes tiling the torus. This will give us the entropy density (\ref{density}).  While the computation of entropy is the same as in the `black hole gas' model of \cite{bf},  the picture in fig.\ref{ffour}  replaces the black holes by states that can be understood in more traditional terms.

 \subsection{Relation to the intersecting brane gas of \cite{cm} }\label{secdiff}
 
 In \cite{cm} an expression was proposed for the entropy in cosmology based on  the above discussed expressions for the entropy of black holes in string theory.   But the expression in \cite{cm} was  different from the expression (\ref{one}) that we have studied in this paper. In this section we note that the difference between these expressions can be traced to a difference in the  assumptions about which states can fit in our box. 
  
  The set up in \cite{cm} was similar to the one here: we take a box of volume $V$,  put an energy $E$ in it, and ask for $S(E,V)$. The conjecture for the entropy, however, was derived by a direct extrapolation of the expressions for black hole entropy in terms of branes and anti-branes. Consider for example the case where string theory is compactified on $T^5$; the entropy of the  black hole in the remaining 4+1 noncompact dimensions is given by   (\ref{threec}) where three types of branes and antibranes are wrapped around the cycles of the compact $T^5$. A similar expression gives the entropy of the hole in 3+1 noncompact dimensions; now the compact directions form a $T^6$ and there are {\it four} sets of intersecting branes and antibranes \cite{4charge}:
\be
S=2\pi (\sqrt{n_5}+\sqrt{{\bar n}_5})(\sqrt{n_1}+\sqrt{{\bar n}_1})(\sqrt{n_p}+\sqrt{{\bar n}_p}) (\sqrt{n_k}+\sqrt{{\bar n}_k}) 
 \label{fourc}
 \ee
  If we compactify all directions, we can wrap even more types of branes on the cycles of the torus. Let the different types of branes (i.e., branes wrapping different sets of cycles) be labelled by $i=1, \dots N$. We take $n_i$ branes and $n_i$ antibranes of each type, so that the configuration is overall neutral. Extrapolating expressions like (\ref{threec}),(\ref{fourc}), it was conjectured in \cite{cm} that the entropy would be\footnote{See \cite{kalyan1} for a detailed analysis of intersecting branes in the early Universe. The consequences of U-duality for the intersecting brane gas were studied in \cite{kalyan1p}. Recently, the behavior of states made from intersecting branes was studied in \cite{rama2}.}
  \be
  S\sim \prod_{i=1}^N (\sqrt{n_i}+\sqrt{{\bar n}_i})
  \label{intersect}
  \ee
  If we take $n_i\sim n$ for all $i$, we have
  \be
  S\sim n^{N\over 2}
  \ee
  Let each brane have $p$ spatial dimensions, and let the  tensions $T_i$ be of order the planck scale. Let the length of each direction of the torus be $L$. Then the total energy is
  \be
  E\sim {L^pn\over l_p^{p+1}}\sim {V^{p\over 9}n\over  l_p^{p+1}}
  \ee
  We find (noting that $G\sim l_p^8$)
  \be
  S\sim {E^{N\over 2} G^{N(p+1)\over 16}\over V^{Np\over 18}}
  \label{st}
  \ee
  This rises more rapidly with $E$ than the expression (\ref{one}). The difference can be traced back to a different choice of assumptions governing the underlying physics of microstates in the large $E$ limit. In (\ref{st}) it is assumed that all states that arise from brane intersections can exist inside the given volume $V$. As was noted in \cite{early}, this need not be the case; when we limit the volume to a given value, not all the states corresponding to the entropy (\ref{intersect}) may be able to exist as orthogonal wavefunctions in this volume. In \cite{phase} it was noted that in the fully solvable case of the 2-charge extremal hole, if we limited the volume available to the states to a value smaller than that set by the Schwarzschild radius, then only a fraction of the full count of states were able to exist. The entropy (\ref{one}) suggested  by duality invariance is less than the entropy (\ref{st}), and so we infer that not all the states corresponding to the entropy (\ref{st}) are able to live in our box of volume $V$ in the density domain (\ref{two}).
  
  This fact can be seen explicitly in the pictorial depiction of fig.\ref{ffour} which reproduces the entropy (\ref{one}). The branes in a given cluster intersect other branes in the same cluster, but not branes in far away clusters. The entropy expression (\ref{intersect}) assumes that each brane of type $i$ intersects {\it all} other branes of type different from $i$. Because of the local nature of brane intersection in fig.\ref{ffour}(a),  the entropy (\ref{one}) is extensive in $V$. The entropy  (\ref{st}), on the other hand is not extensive in $V$. 
 
  \section{The possibility of large quantum effects in cosmology}\label{secsix}
  
 We can put the equation of state (\ref{eos}) in Einstein's equations and find the evolution of the metric of our torus. Note however that 9+1 dimensional string theory contains a dilaton field $\phi$, and the value of this dilation would typically change as the torus size evolves. To avoid this complication it is simplest to work with 11-dimensional M theory, where the size of the extra direction encodes the dilation. Now all the 10 directions of the spatial torus are on a symmetrical footing in Einstein's equations, and there is no additional dilation field. The lengths $L_i\equiv a_i$ of the sides of this M-theory torus are scale factors for our cosmology. The general solution of Einstein's equations with an equation of state of the form $p_i=w_i\rho$ was given in \cite{cm}.  
We have already noted in (\ref{eld}) that our expression (\ref{one}) for the entropy remains unchanged when expressed in M-theory variables, and we have seen in (\ref{weo}) that this expression for the entropy corresponds to   $w_i=1$ for all directions $i$. Let us set all directions to have the same length $L_i\equiv a_i=a$.   Then the Einstein equations give
 \be
({\dot a \over a})^2={8\pi G\over 45} \rho
\label{dthree}
\ee
\be
{\ddot a \over a}=-{8\pi G\over 45}(4 \rho + 5 p)=-{8\pi G\over 5}\rho
\ee
which gives
\be
a\sim t^{1\over 10}, ~~~~\rho\sim {1\over a^{20}}
\label{evolution}
\ee
While the evolution (\ref{evolution})  arises from the classical Einstein equations, it is not clear if these equations should hold as they stand in the present situation. In black holes, it has been observed that quantum effects are large enough to alter the semiclassical physics at the horizon. In this section we recall how these effects arise for black holes, and discuss the possibility of similar effects in cosmology.

\subsection{Large quantum effects in black holes }

In black holes, quantum gravity effects can be large because the black hole has a large degeneracy of states, given by ${\cal N}\sim Exp[S_{bh}]$, where 
\be
S_{bh}\sim {A\over G}\sim {GM^2}
\ee
Here we used the relation $A\sim (GM)^2$ for the 3+1 Schwarzschild hole. The classical action for black hole collapse is
\be
S_{cl}\sim {1\over G}\int {\cal R} \sqrt{-g} d^4 x \sim {1\over G} (GM)^2\sim GM^2
\label{estimate}
\ee
where we have assumed that all length scales are $\sim GM$, and noted that ${\cal R}\sim (GM)^{-2}$. In the path integral
\be
Z\sim \int D[g] e^{-S_{cl}[g]}
\ee
we expect that the measure term is order $Exp[S_{bek}]$. We then see that in the process of gravitational collapse, the measure term competes with the classical action, and semiclassical physics based on $S_{cl}$ alone need not be accurate \cite{tunnel}. More precisely, we find the following possible scenario. In string theory we understand the nature of the $Exp[S_{bek}]$  microstates of the hole; they are given by fuzzballs, whose structure we will discuss below.  Eq. (\ref{estimate}) can be used to estimate the amplitude ${\cal A}$ for the collapsing shell to tunnel into one of the fuzzball microstates:
\be
{\cal A} \sim e^{-S_{cl}}
\ee
While this gives a very small tunneling probability, we must multiply this probability with the large number $Exp[S_{bek}]$ of possible final states. One then finds that the  smallness of the  tunneling probability can be cancelled by the largeness of the degeneracy of final states, and the collapsing shell can change into a linear combination of fuzzball states in a short time. Thus the semiclassical approximation can be violated in the process of gravitational collapse of the shell.

\begin{figure}[h]
\includegraphics[scale=.50]{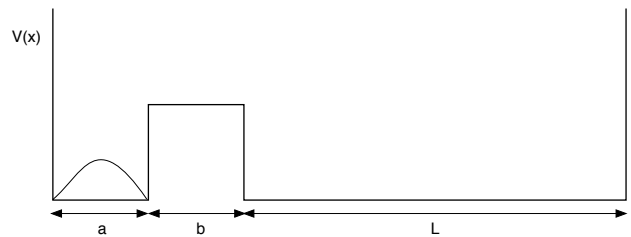}
\caption{\label{ften} A 1-dimensional potential; the particle wave function in the well on the left can tunnel through the barrier into the region on the right.}
\end{figure}

At first it may appear that the above argument is too quick, for the following reason. Consider the process of tunneling in one dimension with the  potential $V(x)$ given in fig.\ref{ften}. There is a potential well of width $a$ on the left, followed by a barrier of height $V_0$ and width $b$, and then  we have a low potential region with large length $L$. A particle in the left well can tunnel into the  dense band of states in the right region. As we let $L$ go to infinity, the density of states in the right region goes to infinity, but the rate of tunneling saturates to a finite value; it does not go to infinity. In fact  the rate of tunneling is set by the height and width of the barrier, and not by the length  $L$ which determines  the density of allowed final states. Thus one might think that in the black hole case it does not help to have the large number $Exp[S_{bek}]$ of final states that the collapsing shell  can tunnel to. 

But in \cite{tunnel} it was argued that tunneling in  the black hole case should be modeled  in a different way from the 1-dimensional  potential $V(x)$. Consider the Hamiltonian for the Schrodinger equation in $d$ dimensions
\be
\hat H=-{\p^2\over \p x_1^2}-{\p^2\over \p x_2^2}-\dots -{\p^2\over \p x_d^2}+ V(x_1)+V(x_2)+\dots V(x_d)
\label{sone}
\ee
where each direction $x_i$ has the same quantum mechanical potential as  the above 1-dimensional  problem. We see that there is a potential well at the center of this $d$ dimensional space, given by the region 
\be
0\le x_i\le a, ~~~i=1, \dots d
\ee
Consider an initial condition where a particle is placed in this central well. In the 1-dimensional potential of fig.\ref{ften}, the probability for the particle to remain in the well decayed with time as $P(t)\sim Exp[-\epsilon \, t]$, where $\epsilon\ll 1$ if we choose the barrier to be tall. In the $d$-dimensional case, the probability for the particle to remain in the central well decays as
\be
P(t)=P_1(t)P_2(t)\dots P_d(t) \sim e^{-d \, \epsilon \, t}
\ee
If we let $d$ be large, so that $d\, \epsilon  \gg 1$, then in a time 
\be
t_{tunnel}\sim (d\, \epsilon )^{-1}\ll 1
\ee
the particle leaves the central well and ends up in a wave function that is a linear combination of states in the regions outside the well. Thus we see that having a large number of different final states to which one can tunnel {\it via different directions} indeed enhances the rate of tunneling.

\begin{figure}[h]
\includegraphics[scale=.60]{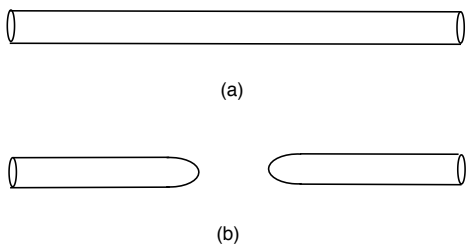}
\caption{\label{fel} (a) Minkowski space-time with an additional compact direction; for simplicity we depict only one spatial noncompact direction. (b) The compact circle can `pinch-off', creating a 'bubble of nothing'.}
\end{figure}

\begin{figure}[h]
\includegraphics[scale=.42]{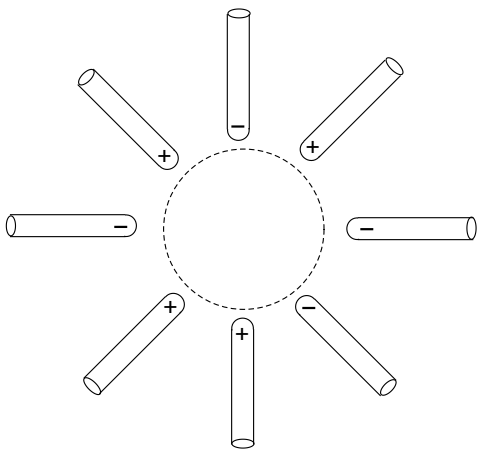}
\caption{\label{ftw} The fuzzball structure of black hole microstates in string theory. A compact direction pinches off with a twist that creates a KK monopole or antimonopole; these two possibilities are denoted by the $\pm$ signs. Spacetime ends just outside the location where the horizon would have formed in the traditional hole. The different choices of monopole structure at different angular locations give the $Exp[S_{bek}]$ micro states of the hole.}
\end{figure}

In \cite{tunnel} it was argued that the black hole case was of this latter type; there are many possible directions in phase space leading to possible fuzzball states, and this corresponds to having a large $d$ in the above toy problem. The fuzzball states can be thought of as eigenstates with mass $M$ for the full string theory Hamiltonian; we may regulate these eigenstates at infinity by putting them in a large box of length $L$. To understand the structure of fuzzballs, we first recall the structure of the `bubble of nothing' that can be formed when we have Minkowski space-time with an additional compact circle \cite{witten}. In fig. \ref{fel} we depict the bubble of nothing in a 1-dimensional illustration.  The compact circle can `pinch-off'; the space-time then ends at this pinch-off radius $R$ and there is no space-time region at $r<R$.  In a fuzzball we have a more complicated pinch-off, where the compact direction twists to make a KK monopole or antimonopole; we denote these two possibilities by `+' and `-' signs respectively in fig.\ref{ftw}. Additional fields and sources in string theory support the monopole structure, so we should just say in general that the space-time ends in a collection of allowed string sources. The different choices of the signs $\pm$ at different angular positions lead to the $Exp[S_{bek}]$ states of the hole. 

When we examine the tunneling paths that lead to the states in fig.\ref{ftw}, we find that there are large number of possible directions to tunnel into, and so the $d$ dimensional model (\ref{sone}) looks more relevant than a 1-dimensional one. With this `multi-directional' tunneling, the rate of tunneling can indeed be very high, and the semiclassical approximation at the horizon can be violated. To write down the evolution of a collapsing shell, we should first write the state of the shell in terms of the fuzzball eigenstates
\be
|\psi_{shell}\rangle= \sum_i C_i |E_i\rangle
\ee
and the subsequent evolution of this state
\be
|\psi_{shell}(t) \rangle= \sum_i C_i e^{-i E_i t} |E_i\rangle
\ee
will show that a shell changes to a linear combination of fuzzball states as it approaches the horizon \cite{rate}. (Thus one need not introduce the `interior' of the hole at all in this computation.) The `effective' dynamics of the hole is given by collective modes that describe simple distortions of the coefficient set $\{ C_i \}$; this is described in more detail in \cite{mt2}. Since this dynamics requires us to consider the amplitudes $C_i$ for all possible configurations (labelled by $i$), we see that the correct way to study the quantum evolution of the black hole is in `superspace' - the space of all possible configurations. 

In black holes the semiclassical approximation is violated because the entropy $S_{bek}$ is large. This largeness stems from the fact that $S_{bek}= A/4G$ has the Newton constant $G$ in the denominator, which brings in the very small  length $l_p$ into the computation  of the entropy. In our present cosmological problem the entropy (\ref{one}) also has a factor $G$ in the denominator. Thus the degeneracy of states in this phase is expected to be high, and we may have quantum effects that invalidate semiclassical dynamics, just as happened for black holes. 

\subsection{Transition into a band: a toy model}

\begin{figure}[h]
\includegraphics[scale=.42]{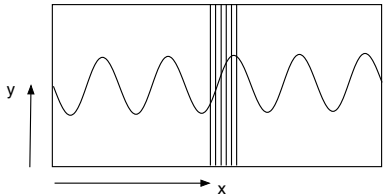}
\caption{\label{ffive} A graviton is placed in a box with wavenumber along the $x$ direction. The box contains strings aligned along the $y$ direction. If the number of strings is large, the graviton is quickly absorbed onto the string as a pair of vibrations.}
\end{figure}

Before addressing the case of cosmology, we consider another toy model where a large degeneracy of states drives the effective evolution.

Consider the situation depicted in fig.\ref{ffive}. We have a graviton $h_{ij}$  in a box, with nonzero wavenumber along the $x$ axis. In the box we have a  a string wound along the $y$ direction, with winding number $N_w$. The string can carry  vibrations moving in the positive and negative $y$ directions, in the form of travelling waves with polarization in any of the directions   transverse to the string.  The full Lagrangian has the form 
\bea
{\cal L} &\sim& \h \p h_{ij}\p h_{ij} + \h \p_+ X^k\p_-X^k\nn
&& \quad +A \, h_{ij}(\p_+X^i \p_-X^j+\p_- X^i\p_+ X^j)
\label{lag}
\eea
where $\p_\pm=\p_t\pm \p_y$, the $X^i$ are the transverse displacements of the string  and $A$ is a constant. This is just the microscopic model used to describe the absorption of gravitons by the D1D5 black hole \cite{callanmalda,dm}, so we may borrow the analysis from that treatment. The classical equations of motion for the excitations on the string give
\be
\p_+\p_- X^i \sim \p_+(h_{ij} \p_- X^j)+\p_-(h_{ij}\p_+ X^j)
\ee
Thus if the initial state has no excitations on the string ($\p_+ X^j=\p_-X^j=0$) then, classically, no excitations will be created on the string. As a consequence the  energy of the graviton will stay in the graviton.  

Quantum mechanically, the situation is different.  The cubic coupling in (\ref{lag}) contains a term of the form $\hat a _h  \, {\hat a}^\dagger _{X_+} {\hat a}^\dagger _{X_-}$ which converts the graviton into a pair of oppositely moving excitations on the string. Since this pair creation is a quantum effect, it would normally be considered small. But the situation changes if $N_w$ is very large, as is the case in the effective string model of the D1D5 black hole. The density of energy levels on the string is
\be
\rho_E\sim {N_w\over L_y}
\ee
where $L_y$ is the length of the $y$ direction along which the string is wrapped. The rate of absorption of the graviton onto the string is proportional to $\rho_E$, and for large $\rho_E$, is quite quick. 

When the energy resides in the graviton, we find a pressure $p_x>0$ in the $x$ direction since the graviton was chosen to have a wavenumber in the $x$ direction. With such a pressure, if we allowed the walls of the box to expand, they would expand in the $x$ direction. After the graviton is absorbed onto the string, we find $p_x=0$ but a  pressure $p_y$ in the $y$ direction.\footnote{This pressure arises because the energy $E_X$ of a vibration mode in the $n$th harmonic on the string behaves as $E_X\sim {n\over N_w L_y}$. Since this energy is higher for smaller $L_y$, the vibrations cause a positive pressure $p_y>0$.} If we allowed the box to expand in response to this pressure, it would expand along the $y$ direction.

To summarize, if a system can access a set of states with very high level density $\rho_E$, then its state can move into the band of such states through a `fermi golden rule absorption' at a rate proportional to $\rho_E$.  As a consequence the system can evolve in a  manner different from that expected from the classical equations of motion. 

\subsection{A scenario for evolution in the  early Universe}\label{secsc}

Let us now ask how it may be possible to violate the classical evolution (\ref{evolution}) because of the large entropy (\ref{one}). A scenario for this violation was discussed in \cite{early}. This scenario uses the fact noted in section \ref{secdiff} that in our chosen phase, not all states of energy $E$ are able to fit in the volume $V$. If the volume $V$ were to quantum-fluctuate to a larger value, then many more states would be able to fit in, while if it quantum-fluctuated to a smaller value, then many fewer states would fit in. It was then argued that this circumstance drives an expansion to larger scale factors, above and beyond any expansion that may result from the classical Einstein equations.\footnote{See also \cite{kalyan2} for a discussion of entropy effects in the early Universe.}

\begin{figure}[htbp]
\begin{center}
\includegraphics[scale=.48]{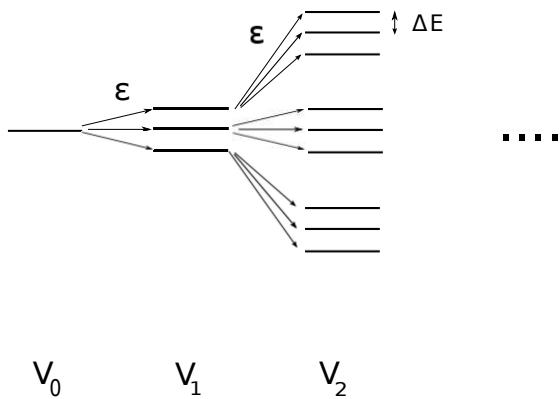}
\caption{{The states at each size $V_k$ can transition with amplitude $\epsilon$ to a band of states in volume $V_{k+1}$, with band spacing $\Delta E$; thus we get a series of `fermi golden rule absorptions' taking us to larger volumes.}}
\label{fsix}
\end{center}
\end{figure}

Such an effect would of course be present in many quantum systems, but the circumstance which makes it interesting in our present cosmological problem is that the number of states involved is very large; as we had noted above, this largeness stems from the appearance of $G$ in the denominator of (\ref{one}) which brings in the planck scale.  For every state available at one scale factor, there is a densely spaced band of states into which it can evolve if the scale factor were larger. The quantum mechanical problem describing this situation is pictured in fig.\ref{fsix}.  We consider a sequence of volumes for a spatial box, starting at a volume $V_0$, and moving through progressively larger volumes $V_1, V_2, \dots$. The state in the box of volume $V_0$ can be absorbed into a band of states in the volume $V_1$.  But each state in this band can itself be absorbed into a band of states in the larger volume $V_2$ and so on. 

For our cosmological situation we do not know the amplitudes for the transitions between levels, but we make a toy model by seting all amplitudes for transition per unit time  to be equal to the same number   $\epsilon$. We also let the energy spacing in each band be the same, $\Delta E$.  As the system evolves, the wavefunction moves from the state at volume $V_0$ to the band of states in the volume $V_1$, then to the states at $V_2$ and so on. We can compute  the probability to be at volume $V_k$ aftertime $t$.  In particular, we can ask for the value of $k$ where this probability peaks at any given time $t$. One finds that the location of this peak is given by \cite{early}
\be
k_{peak}(t)\approx {2\pi \epsilon^2\over \Delta E}t
\ee 
Note that  this expansion $V_0\rightarrow V_1 \rightarrow \dots $ is in {\it addition} to any expansion rate obtained from the classical gravity equations, since it is generated by the phase space measure describing the degeneracy of states. This measure factor is small in typical laboratory systems, but can be very large if the entropy has a gravitational origin involving the Newton constant $G$. 

This evolution to larger volumes $V_k$ may be quite rapid, and may give a kind of `inflation' without the need for an inflaton field with a slow roll potential. Initial density fluctuations are likely to arise from an effective action derived from the free energy, just as was found for the string gas \cite{bvspec}.\footnote{The evolution of perturbations in  a `black hole gas' was studied in  \cite{bf}.} These fluctuations are also likely to stay frozen in amplitude as the volume evolves through the steps in fig.\ref{fsix}, for the following reason.  Suppose  the local energy density at a point is a little higher than the average ($\rho=\rho_{av}(1+\delta)$). Then this higher value of the local energy stays fixed as the system moves through the steps of fig.\ref{fsix}; it does not dissipate away to neighbouring lower density points because the evolution of fig.\ref{fsix} does not have a kinetic term that moves energy from one point to neighboring points. fluctuations. If we do get a rapid expansion with perturbations that are frozen in amplitude, then we mimic the results of an inflationary scenario. When the branes of fig.\ref{ffour} finally annihilate to radiation, we would get these perturbations imprinted onto that radiation.

\section{Discussion}\label{secseven}

It is interesting that one can reach the entropy expression (\ref{one})  from  very different arguments. In \cite{sas1} this expression was obtained from a `spacetime uncertainty relation'. In  \cite{fs,bf} one considers the expansion caused by the matter density $\rho$, and assumes that the entropy in a cosmological horizon region would be of order the Bekenstein entropy for the largest hole that would fit in this region. The third line of argument (noted in \cite{sas} and analyzed in detail in the present paper)  requires that the expression for entropy be invariant under the T and S dualities of string theory. This requirement limits the possible functional dependences for $S(E,V)$. In \cite{sas} it was argued that we get (\ref{one}) if we further require that $S$ be proportional to $V$. In our analysis we did not imposed this requirement, but  requiring agreement with the Bekenstein entropy for $E\sim E_{bh}$ gave the expression (\ref{one}).  

We have noted that  while T and S dualities are exact symmetries of string theory, it does not follow that the expression for the entropy $S$ must be invariant under these dualities. The {\it value} of the entropy $S$ will necessarily be invariant, since the dualities are a symmetry. But the {\it expression} for $S$ need not be invariant; it will in general only be covariant, changing form as we change duality frames. Thus the requirement that the expression for $S$ be invariant is an additional assumption, and it is this assumption, coupled with the requirement that $S\sim S_{bek}$ at $E\sim E_{bh}$ that gave us the entropy expression (\ref{one}). The requirement that the expression for $S$ be invariant under dualities is suggested by the invariances found in the expressions for the entropy  of black holes in string theory.

In \cite{bf} it had been observed that an entropy like  (\ref{one}) can be obtained by taking a densely spaced set of black holes, with the radius of the holes being of order the Hubble radius. Black holes might appear to be esoteric objects, but we have learnt in string theory that their entropy can be reproduced by counting the states of intersecting branes. In \cite{veneziano} it was noted that when the energy density $\rho$ is string scale, a dense gas of string states at the `Horowitz-Polchinski correspondence point' reproduces the entropy (\ref{one}). We have noted that at any density $\rho$ we can replace the black holes by sets of intersecting branes, so an entropy like (\ref{one}) can be realized by states in string theory. These intersecting brane states do not collapse into black holes; instead they generate `fuzzball states' which are complicated states of string theory without horizons or singularities. 

Finally, we explored the idea that the evolution of such a high entropy state may not satisfy the traditional Einstein equations. The entropy (\ref{one}) matches onto the black hole entropy when $E\sim E_{bh}$. In black holes we get a traditional horizon if we assume that Einstein's equations are satisfied by a  shell as it collapses through its horizon. But the horizon so generated leads to the black hole information paradox, which is a serious obstruction to the unitarity of the  underlying quantum theory. In string theory the semiclassical approximation can be violated at the horizon because the collapsing shell can tunnel into a densely spaced band of fuzzball states. We can therefore ask if  similar effects can come into play in our cosmological situation.  We have noted that the number of states increases rapidly with the volume $V$; this rapid increase can be traced to the appearance of $G\sim l_p^{d-1}$ in the denominator of (\ref{one}) which makes $S$ very large. If the density of state rises very rapidly with the volume $V$, then the toy model constructed in section \ref{secsc} suggests that there may be a  rapid `push' towards larger $V$. This push would arise from the measure in the path integral (which tracks the number of available states) and would therefore be in addition to any expansion arising from the classical Einstein equations. This argument is certainly speculative, but it would be interesting to study it further. 
    
\begin{acknowledgments}
We thank  R. Brandenberger, R. Bousso, S. Das, S. Kalyana Rama, J. Maldacena, A. Sen, D. Turton,  E. Verlinde and H. Verlinde for comments.
This work was supported in part by DOE grant DE-FG02-91ER-40690.
\end{acknowledgments}

\nocite{*}


\end{document}